\documentclass[twocolumn,showpacs,aps]{revtex4}
\usepackage[dvips]{graphicx}

\usepackage{times}
\begin{document}

%\begin{CJK*}{GBK}{song}
\title{Enhancing parameter precision of optimal quantum estimation by direct quantum feedback}
%\thanks{Supported by the National Natural Science Foundation of
%China under Grant Nos. 11065005 and 11365006.
%\\*** Email: qzhengnju@gmail.com
% Tel:152-8601-8266; 182-8619-3565;
% Email: qz@gznu.edu.cn}

\author
{
Qiang Zheng $^{1,~2}$, Li Ge $^{2}$, Yao Yao $^{2}$, and Qi-jun Zhi $^{3}$
%\footnote{ Email: liyong@csrc.ac.cn}
}
\address{
$^{1}$ School of Mathematics and Computer Science, Guizhou Normal
University, Guiyang, 550001, China
\\
$^{2}$ Laboratory for Quantum Optics and Quantum Information, Beijing Computational Science Research Center, Beijing 100084, China
\\
$^{3}$ School of Physics and Electronics, Guizhou Normal
University, Guiyang, 550001, China
}

\begin{abstract}
Various schemes have been proposed to overcome the drawback
of the decoherence on quantum-enhanced parameter estimation.
Here we suggest an alternative method, quantum feedback,
to enhance the parameter precision of optimal quantum estimation of a dissipative qubit by investigating its dynamics of quantum Fisher information. We find that compared with the case without feedback, the quantum Fisher information of the dissipative qubit in the case of feedback has a large maximum value in time evolution and a smaller decay rate in the long time.
\end{abstract}

%\date{today}
\pacs{03.67.-a; 06.20.-f; 02.30.Yy;}

\maketitle

\section{Introduction}

Quantum metrology~\cite{VGio11}, which investigates the limit of the precision of parameter estimation bounded by the quantum mechanics, excites wide interest in recent years. One of the basic ideas is to surpass the shot-noise limit
by making use of non-classical resources which provide, e.g., NOON state or entangled state. Quantum metrology has wide application in improving time and frequency standards~\cite{Udem02, Bollinger, Huelga07}, detecting gravitational
waves~\cite{Aasi2}, and magnetometry~\cite{Taylor08}, and so on.

Quantum Fisher information (QFI) plays a central role in quantum metrology and quantum estimation theory. According to quantum Cram\'{e}r-Rao inequality, the QFI imposes an upper bound on the precision of parameter estimation. A larger QFI implies that the parameter can be estimated with a higher precision. In addition to the basic properties such as convexity and monotonicity, the QFI is also closely connected with other quantities, especially entanglement~\cite{Smerzi09}, non-Markovianity~\cite{xmLu10}, and spin squeezing~\cite{Strobel14,jma11}. More importantly, the QFI closely connects with other branches of quantum physics, such as quantum phase transition~\cite{Salvatori14}, quantum clone~\cite{yaoyao14}, and quantum Zeno dynamics~\cite{Smerzi12}.

% decoherence and feedback
Each realistic quantum system inevitably interacts with its environment,
which leads to decoherence and considerable decrease of precision of parameter estimation~\cite{Escher11, Guta12, zyou14, jin13, wangx11, qz14}.
In order to enhance QFI, various methods, such as quantum error correction, dynamical decoupling, decoherence-free subspace, reservoir engineering, have been proposed to suppress the decoherence. For instance, an approach based on decoherence-free subspace has been proposed~\cite{Dorner12} to enhance the precision of estimation of atomic transition frequencies with ions stored in Paul traps subject to collective dephasing. Adopting the method of dynamical decoupling, the maximum Fisher information of a single qubit in a noisy system has been retrieved~\cite{Ueda10}. Inspired by Ref.~\cite{Ueda10}, Tan~\textit{et~al.} proposed a scheme to enhance the QFI of a qubit by employing dynamical decoupling pulses~\cite{tan13}. Berrada~\cite{Berrada} and Chin~\textit{et~al.}~\cite{Chin12} also considered the QFI of a qubit subject to non-Markovian environment, which can be viewed as an engineered reservoir.

Besides the above mentioned strategies, it is also interesting to use alternative methods to enhance QFI. Here we propose using the quantum feedback to enhance the precision of parameter estimation of a dissipative qubit. Quantum feedback~\cite{Wiseman93, Wiseman95} has been considered as one of the essential ways to suppress the decoherence, which manipulates the system based on the information acquired by measurement. Compared with the other control approaches, it operates simply by feeding back some information, which is proportional to the measurement signal, directly to the controlled system~\cite{Wiseman07}. Generally, it is well known that the feedback can only partly compensate the decoherence effect of an open system governed by the Lindblad master equation. In the method of direct feedback, there are various schemes to measure the output field, such as homodyne or direct photodetection. We will adopt the latter scheme and only consider the perfect detection efficiency in this paper.

As a first step to explore the possibility that quantum feedback
can beat the shot-noise limit and eventually achieve the Heisenberg limit,
we study a qubit system. Moreover,
motivated by recent experimental achievement on quantum feedback control on a qubit system~\cite{DiCarlo13, Vijay12}, in this paper we study the parameter estimation of a dissipative qubit in a cavity which provides the quantum feedback to the qubit. For the purpose of exploring the effect of quantum feedback, we only focus on the QFI of the qubit with considering the optimal conditions for saturated quantum Cram\'{e}r-Rao inequality.
%In the bad-cavity limit, the cavity mode can be adiabatically eliminated, and the qubit also obtains an %effective damping rate.
%We study the QFI of the qubit with respect to the damping rate.
Under the effect of quantum feedback, we find that: %(i) in the short-time evolution, the QFI with feedback is smaller than that without feedback;
(i) in the long-time limit, the decay rate of the QFI with feedback is slowed down; and (ii) the maximum value of the QFI in the evolution is also enhanced. These results imply that the quantum feedback can enhance the precision of parameter estimation. In addition, we find there is an optimal external driving strength for the qubit maximizing the QFI of the steady state.

This paper is organized as follows. The basic properties of the QFI are reviewed in Sec.~II. In Sec.~III, we investigate the feedback effect on the QFI of a qubit interacting with a cavity, and find that the QFI is improved by feedback and discuss the behaviours of the QFI in the steady state with the external driving of the qubit. Finally, a summary is provided in the last section.

\section{Quantum Fisher information}
In this section, we will review the main aspects of QFI. Let $\varphi$ denote a single parameter to be estimated, and $p_{i}(\varphi)$ be the probability density with measurement outcome $\{ x_{i} \}$ for a discrete observable $X$ conditioned on the fixed parameter $\varphi$. The classical Cram\'{e}r-Rao inequality~\cite{Holevo} gives the bound of the variance $\mathrm{Var}(\hat{\varphi})$
for an unbiased estimator $\hat{\varphi}$
\begin{equation}
\begin{array}{llll}
\mathrm{Var}(\hat{\varphi}) \geq \frac{1}{H_{\varphi}},
\end{array}
\end{equation}
where the classical Fisher information is defined as \cite{Fisher}
$H_{\varphi}= \sum_{i}p_{i}(\varphi)[\frac{\partial}{\partial \varphi} \ln p_{i}(\varphi)]^2$.

Extending to quantum regime, in order to determine the ultimate bound to precision posed by quantum mechanics, the Fisher information must be maximized over all possible measurements~\cite{paris09}. By introducing the symmetric logarithmic derivative $L_{\varphi}$ determined by
$ \frac{ \partial \rho_{\varphi} }{ \partial \varphi }=\frac{1}{2} (\rho_{\varphi}
L_{\varphi}+ L_{\varphi} \rho_{\varphi} )$,
the so-called quantum Cram\'{e}r-Rao inequality gives a bound to the variance of any unbiased estimator~\cite{Caves94}:
\begin{equation}
\begin{array}{llll}
\mathrm{Var}(\hat{\varphi}) \geq \frac{1}{H_{\varphi}} \geq \frac{1}{F_{\varphi}}.
\end{array}\label{crineqB}
\end{equation}
It is worth mentioning that the symmetric logarithmic derivative $L_{\varphi}$ provides an optimal measurement, that is, the inequality in Eq.~(\ref{crineqB}) can be saturated via using a measurement composed by the eigenvectors of $L_{\varphi}$. Here, the QFI of a quantum state $\rho_{\varphi}$ with respect to the parameter $\varphi$ is defined as~\cite{Caves96}
\begin{equation}
\begin{array}{llll}
F_{\varphi}= \mathrm{Tr} ( \rho_{\varphi} L_{\varphi}^2 ).
\end{array}
\end{equation}
With the spectrum decomposition $\rho_{\varphi}= \sum_{k} \lambda_{k}|k\rangle \langle k|$, the QFI is completely determined by
\begin{eqnarray}
%\begin{array}{llll}
F_{\varphi} &=& \sum_{k, \lambda_{k}>0} \frac{(\partial_{\varphi }\lambda_{k})^2}{\lambda_{k}} \nonumber \\
&+&\sum_{k,~k', \lambda_{k}+\lambda_{k'}>0} \frac{2(\lambda_{k}-\lambda_{k'})^2}{\lambda_{k}+\lambda_{k'}} |\langle k | \partial_{\varphi}k' \rangle|^2.
%\end{array} \label{FisherAA}
\end{eqnarray}
The first term in this equation is just the classical Fisher information,
and the second term can be considered as the quantum contribution. And the
symmetry logarithmic derivative can be obtained as
\begin{equation}
L_{\varphi}= \sum_{k,~k',~\lambda_{k}+\lambda_{k'}>0} \frac{2 \langle k | \partial_{\varphi}\rho |k' \rangle}{\lambda_{k}+\lambda_{k'}}
|k \rangle \langle k'|.
\end{equation}

Usually it is difficult to give the QFI in the form the matrix density for a general system. Fortunately, the QFI of the two-dimensional density matrix has obtained explicitly in Refs.~\cite{Dittmann99, wzhong13} as
\begin{equation}
\begin{array}{llll}
F_{\varphi}= \mathrm{Tr}[(\partial_{\varphi} \rho)^2]+\frac{1}{\mathrm{ Det(\rho)}} \mathrm{Tr}[(\rho \partial_{\varphi} \rho)^2].
\end{array}\label{QFIdef}
\end{equation}

Moreover, the QFI is also related to the Bures distance~\cite{Caves94} through
\begin{equation}
D_{B}^2[\rho_{\varphi}, \rho_{\varphi+d\varphi}]= \frac{1}{4}F_{\varphi}d \varphi^2,
\end{equation}
where the Bures distance which measures the distance between two quantum states $\rho$ and $\sigma$ is defined as~\cite{Nielsen00} $ D_{B}[\rho, \sigma]=[ 2(1-\mathrm{Tr}\sqrt{\rho^{1/2} \sigma \rho^{1/2}})]^{1/2}$.

\section{Enhancing QFI of a qubit with feedback}
\subsection{Model}

\begin{figure}[!tb]
\centering
\includegraphics[width=2.0in]{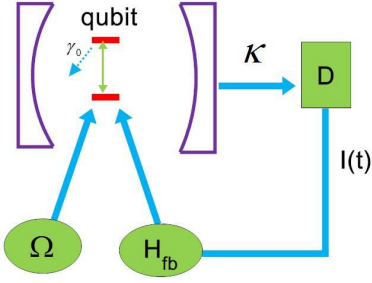}
\hspace{2.5cm}
\caption{(Color online) Schematic diagram of the parameter estimation with feedback.
The qubit is coupled to the cavity resonantly. The output of the cavity is measured,
and the measurement is used to trigger a feedback Hamiltonian. }
\label{system}
\end{figure}

We consider a system consisting of a qubit resonantly interacting with a single-mode cavity, as shown in Fig.~$\ref{system}$. The (real) coupling strength between the qubit and the cavity is $g$, the cavity mode $a$ is damped at rate $\kappa$, the two levels of the qubit are $|e\rangle$ and $|g\rangle$, and its spontaneous emission rate of the excited state is $\gamma_{0}$. Without feedback, the time evolution of the whole system is described by the Lindblad master equation ($\hbar=1$)
\begin{equation}
\begin{array}{lll}
\frac{d \rho}{d t}=-i[H, \rho]+ \kappa \mathcal{D}[a]\rho+ \gamma_{0}\mathcal{D}[\sigma_{-}]\rho ,
\end{array} \label{liouv}
\end{equation}
where the superoperator is defined as
$\mathcal{D}[c]\rho \equiv c\rho c^{\dag}-\frac{1}{2}(c^{\dag} c \rho+ \rho c^{\dag} c )$, and
the Hamiltonian of the model under consideration is given as
\begin{equation}
\begin{array}{llll}
H=\frac{\Omega }{2} \sigma_{x}+g(\sigma_{+}a+ h.c.).
\end{array}\label{Hamit}
\end{equation}
Here we have assumed that the qubit is subject to an external classical drive with strength $\Omega$. The Pauli operators $\sigma_{z}= |e\rangle \langle e|- |g\rangle \langle g|$ and $\sigma_{x}=\sigma_{-}+\sigma_{+}$ with $\sigma_{-}= |g\rangle \langle e|$ and $\sigma_{+}= |e\rangle \langle g|$ being the lowering and raising operators of the qubit, respectively.

Under the condition that the cavity decay $\kappa$ is much larger than the other relevant frequencies
of the system, the cavity mode can be adiabatically
eliminated and the effective damping rate of the qubit is
obtained as~\cite{jwhw05}
\begin{equation}
\begin{array}{llll}
\gamma=g^2/\kappa.
\end{array}\label{gammaA}
\end{equation}
In what follows, we will consider the limit $\gamma \gg \gamma_{0}$ such that the spontaneous emission of the qubit can be omitted.

%The main idea
Now let us consider the system subject to the feedback, as shown in Fig.~$\ref{system}$: The output from the cavity is measured by a detector $D$, then the signal $I(t)$ from the detector $D$ triggers the control Hamiltonian $H_{\rm{fb}}=I(t)B$. For a feedback scheme based on photodetection measurement,
the unconditional state of the qubit is described by~\cite{hope07, li08}
\begin{equation}
\frac{d \rho}{d t}=-i\frac{1}{2}\Omega [\sigma_{x}, \rho]+ \gamma \mathcal{D}[U\sigma_{-}]\rho. \label{liouv2}
\end{equation}
Rewriting the last term of Eq.~(\ref{liouv2}) explicitly:
$\mathcal{D}[U\sigma_{-}]\rho=U \sigma_{-}\rho \sigma_{+} U^{\dag}-\frac{1}{2}(\sigma_{+} \sigma_{-} \rho+ \rho \sigma_{+} \sigma_{-})$, the effect of the feedback is obvious.
Conditioned on a detection event $\sigma_{-} \rho \sigma_{+}$, the unitary evolution $U=e^{-iB \delta t}$ is immediately applied on the system lasting finite time $\delta t$. Throughout this paper, the other relevant parameters are scaled by $\delta t$, such as that $t$ or $\gamma$ denotes the scaled time $\frac{t}{\delta t}$ or decay rate $\gamma \delta t$, and we set $\delta t=1$. Taking the constraint $UU^{\dag}=1$ into consideration (which is required by experiments), we choose the feedback operator as~\cite{xxyi09}
\begin{equation}
\begin{array}{llll}
U= e^{i \overrightarrow{\sigma} \cdot \overrightarrow{A}}=\cos(A) I_{2}+i\sin(A) [\sigma_{x} \sin(\beta)+\sigma_{y} \cos(\beta)].
\end{array}\label{feedoperator}
\end{equation}
Here $I_{2}$ is $2\times2$ identity operator, $\overrightarrow{\sigma}=(\sigma_x, \sigma_y,\sigma_z)$ and $\overrightarrow{A}=(A\sin(\beta), A\cos(\beta), 0)$.
The two parameters $A$ and $\beta$ fully characterize the feedback operator $U$.

%%%%%%%%%%%%%%%%%%%%%%%%
%%%% Figure 1:
%%%%%%%%%%%%%%%%%%%%%%%%
%%%%%%%%%%%%%%%%%%%%%%%%
%%%%%%%%%%%%%%%%%%%%%%%%

\subsection{Without feedback}

In this subsection, we investigate the dynamics of the QFI in absence of the feedback. Here we focus on that the external driving is switched off, the case $\Omega \neq 0$ will be discussed in Sec. III D. The effective damping in Eq.~(\ref{liouv2}) can be considered as a quantum channel for the qubit. Then we adopt the quantum parameter estimation theory to estimate the effective damping parameter $\gamma$.

For an superposition initial state $|\psi_{0} \rangle=
(\cos(\vartheta)|e\rangle+ \sin(\vartheta)|g\rangle)/$, we first
concentrate on $\vartheta=\pi/4$. The effect of population on
$|e\rangle$ will be discussed later. For this equal-weighted
superposition initial state, the evolved density matrix of the qubit
can be exactly solved, which is given as
\begin{equation}
\rho(t)=\left(                 %左括号
\begin{array}{cc}   %该矩阵一共3列，每一列都居中放置
 \rho_{11}(t) & \rho_{12}(t) \\  %第一行元素
 \rho_{12}^{*}(t) &  1-\rho_{11}(t) \\  % 第二行元素
 \end{array}
 \right) \label{densityA}
\end{equation}
with the elements
\begin{equation}
\begin{array}{llll}
 \rho_{11,~off}(t)= e^{-t \gamma}/2,~~\rho_{12,~off}(t)= e^{-t \gamma/2}/2,
 \end{array}\label{densityA2}
\end{equation}
under the condition without the feedback, i.e., by setting $A=\pi$.

Adopting Eq.~(\ref{QFIdef}) by some straightforward calculations, the QFI of $\rho(t)$ in Eq.~(\ref{densityA}) with respect to $\gamma$ is obtained as
\begin{equation}
\begin{array}{llll}
F_{\gamma}(t)= \frac{t^2 e^{-t\gamma}(2 e^{t\gamma}-1)}{4(e^{t\gamma}-1)}.
\end{array} \label{fisherNF}
\end{equation}
%Some interesting limits of Eq.~(\ref{fisherNF}) are considered.
In the limit $t\rightarrow 0$, we find that
\begin{equation}
\begin{array}{llll}
F_{\gamma}(t) \simeq \frac{t}{4 \gamma}+ \frac{t^2}{8}.
\end{array} \label{fishlimit}
\end{equation}
This implies that the QFI increases with time in the initial evolution.
On the other hand, in the long time limit $t \rightarrow \infty$, the QFI reduces to
\begin{equation}
\begin{array}{llll}
F_{\gamma}(t) \rightarrow t^2 e^{-\gamma_{F} t}/2,
\end{array} \label{fishlimitE}
\end{equation}
which exponentially decays to zero with rate $\gamma_{F}=\gamma$.
The similar result has also been obtained in Ref.~\cite{wzhong13},
although a different estimated parameter was considered in Ref.~\cite{wzhong13}, i.e., a phase
parameter within input states, which experience amplitude damping or
spontaneous emission before measurement.

%%%%%%%%%%%%%%%%%%%%%%%%
%%%% Figure 2:
%%%%%%%%%%%%%%%%%%%%%%%%
\begin{figure}[!tb]
\centering
\includegraphics[width=2.0in]{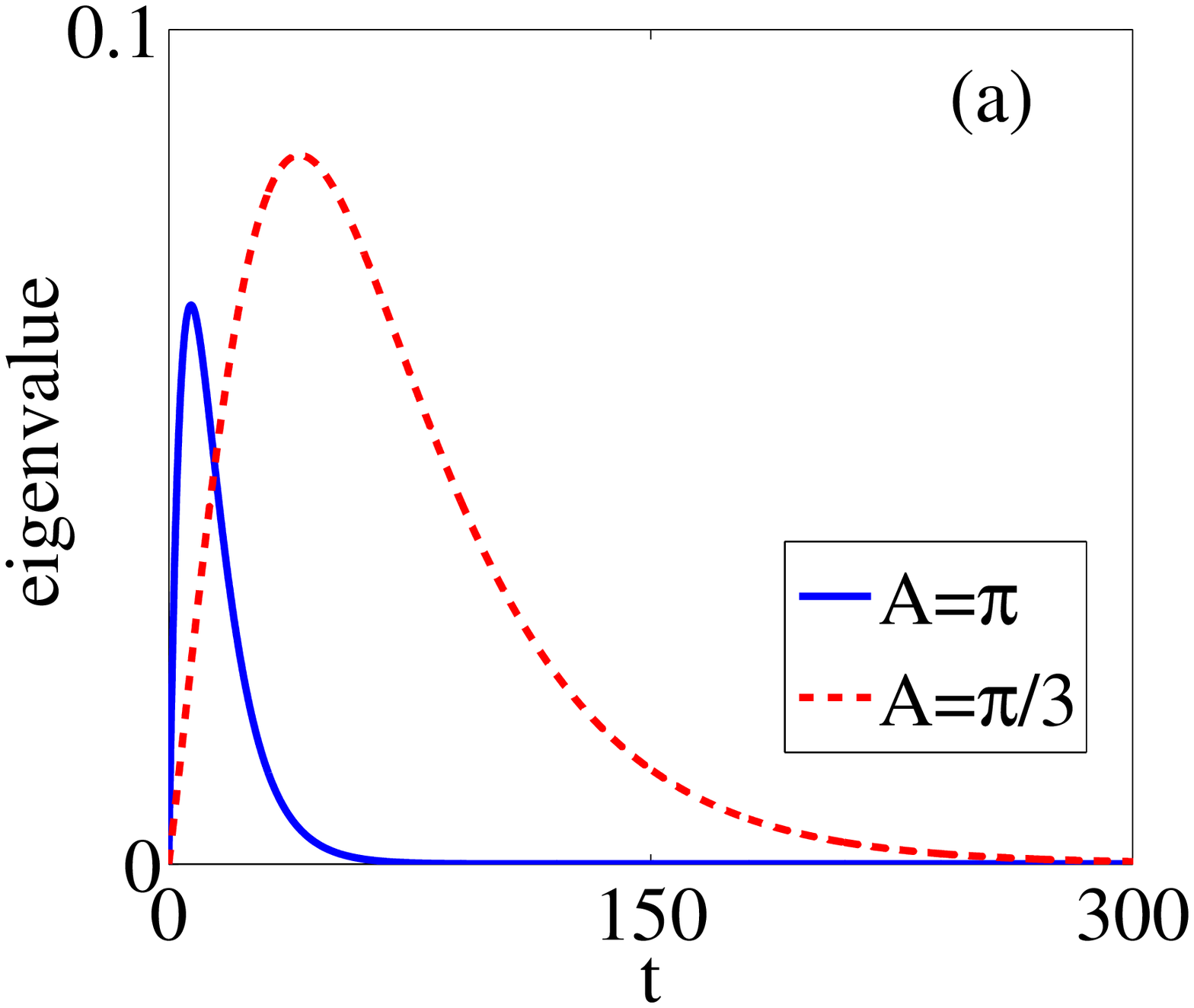}
\includegraphics[width=2.0in]{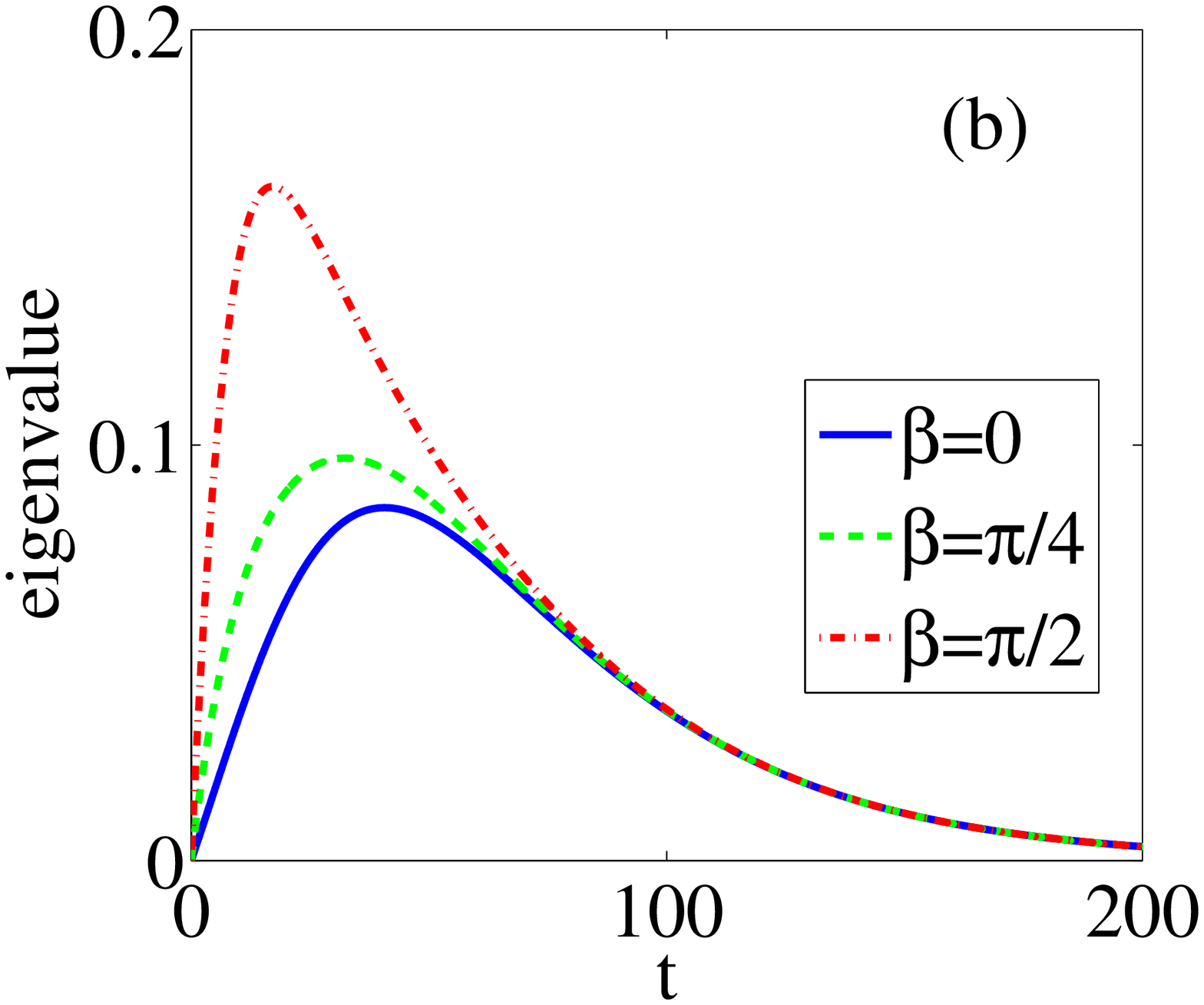}
\hspace{2.2cm}\caption{(Color online) The evolution of the eigenvalue of the qubit density matrix
under the effect of the control parameters $A$ in panel (a) and $\beta$ in panel (b).
Here $\beta=0$ in (a) and $A=\pi/3$ in (b), respectively. The other parameter is $\gamma=0.1$, and $A=\pi$ corresponds to turning off the quantum feedback. All the relevant quantity, scaled by
the feedback lasting time $\delta t$, is dimensionless mentioned here and in the other figures (setting $\delta t=1$).} \label{matfeedA}
\end{figure}
%%%%%%%%%%%%%%%%%%%%%%%%
%%%%%%%%%%%%%%%%%%%%%%%%

\subsection{Enhancing QFI by feedback}

In order to show the effect of quantum feedback, in Fig.~\ref{matfeedA} we plot the evolution of the smaller eigenvalue of the qubit density matrix $\rho(t)$ under the effect of the feedback parameters. Fig.~\ref{matfeedA}(a) shows that the (smaller) eigenvalue of the qubit is increased by the feedback and has
 the slow decay in the long-time limit with $A=\pi/3$.
Fig.~\ref{matfeedA}(b) presents that the eigenvalue decreases with the increase of
$\beta$ in the initial time and
is almost independent of the feedback parameter $\beta$ in the long-time limit.
Thus, we keep $\beta=0$ in the following discussions as a good approximation.

When the feedback is turned on (but still in absence of qubit driving), the evolved elements of the qubit density matrix are given as
\begin{equation}
\begin{array}{llll}
 \rho_{11}(t)= e^{-t \gamma_{q}}/2,~~
 \rho_{12}(t)= \rho_{12,~off}(1+Y),
 \end{array}\label{rhofeed}
\end{equation}
resulting from Eq.~(\ref{liouv2}) with $\beta=0$. Here, it is apparent that
\begin{equation}
\begin{array}{llll}
\gamma_{q}=\gamma \cos(A)^2,~~Y=(1-e^{-t\gamma \cos(2A)/2 }) \tan(2A),
 \end{array}
\end{equation}
are attenuating factors of qubit density matrix induced by the quantum feedback.

%%%%%%%%%%%%%%%%%%%%%%%%
%%%% Figure 3:
%%%%%%%%%%%%%%%%%%%%%%%%
\begin{figure}[!tb]
\centering
\includegraphics[width=2.0in]{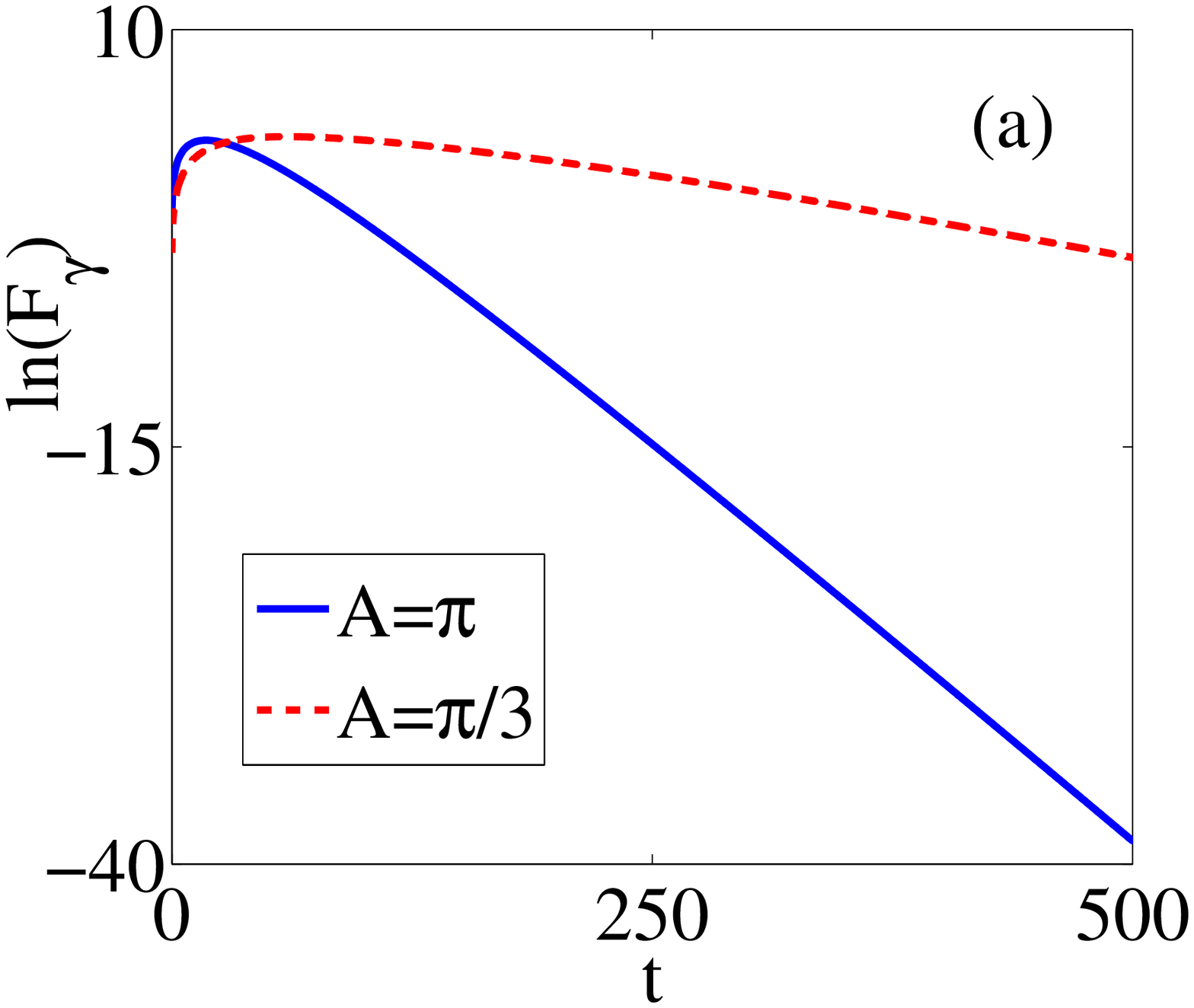}
\includegraphics[width=2.0in]{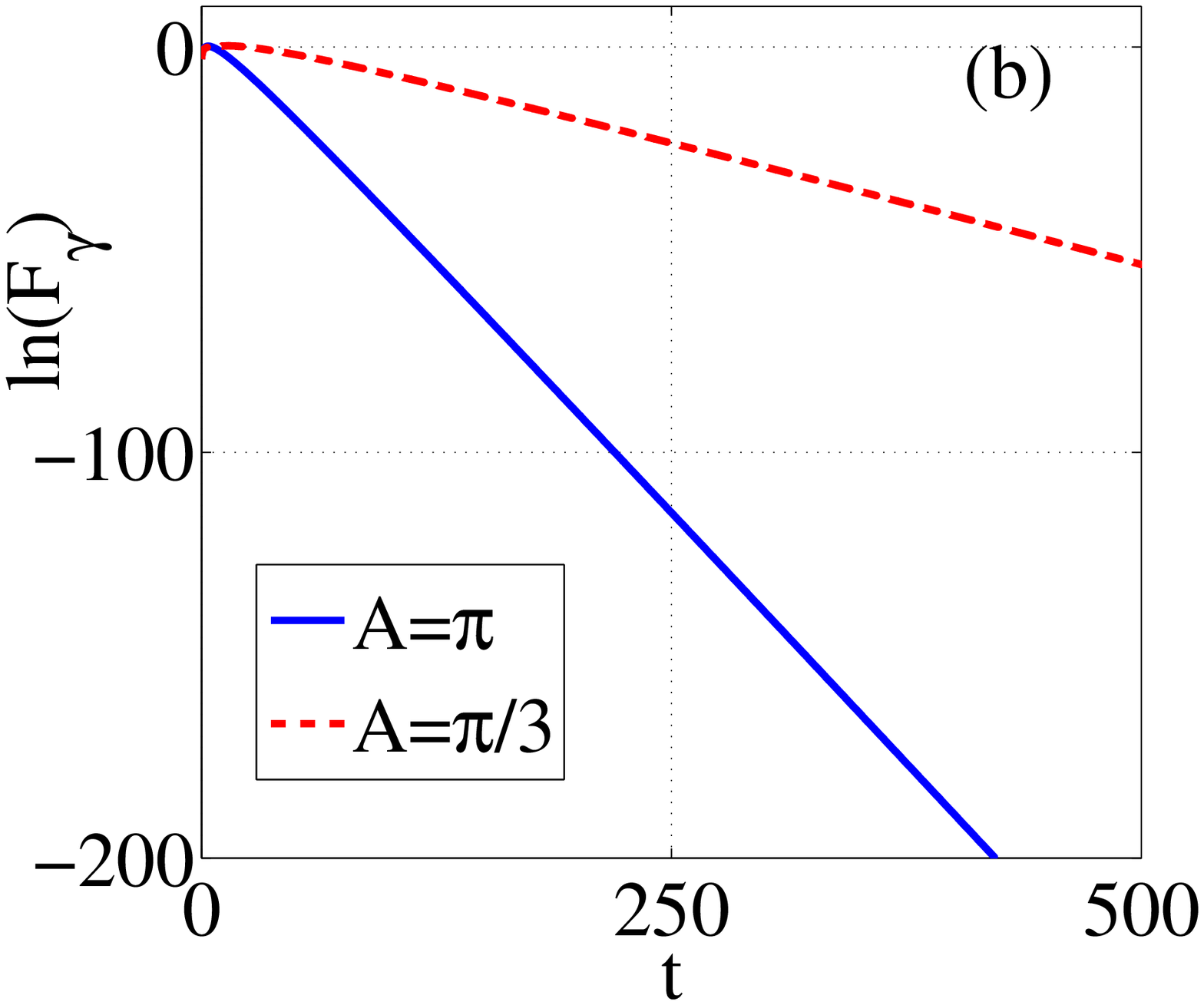}
\hspace{2.2cm}\caption{(Color online) The evolution of the QFI $F_{\gamma}(t)$
with different decay rates $\gamma=0.1$ (a) and $\gamma=0.5$ (b).
The dashed line ($A=\pi/3$) correspond to turning on the feedback, while the solid line ($A=\pi$)
to off the feedback, respectively. } \label{FisherA}
\end{figure}
%%%%%%%%%%%%%%%%%%%%%%%%
%%%%%%%%%%%%%%%%%%%%%%%%

For general feedback parameter $A$, the analytical expression of the QFI is
very cumbersome. As a warm-up, we study two relatively simple situations.
For $A=\pi/2$, the effect of the feedback is equivalent to qubit dephasing:
only the off-diagonal elements decays exponentially with a rate $\gamma_{q}=\gamma/2$.
In this case, the QFI is obtained as
\begin{equation}
\begin{array}{llll}
F_{\gamma}(t)= \frac{t^2}{4(e^{t\gamma}-1)},
 \end{array}
\end{equation}
which also decays exponentially to zero with rate $\gamma_{F}=\gamma$ in the long-time limit. Therefore, there should exist an optimal time to maximize the value of the QFI. For $A=\pi/3$, which makes all the elements of the qubit density matrix in Eq.~(\ref{rhofeed}) exponentially decay to zero with rates $\gamma_{q}=\gamma/4$.
In the long-time limit, the QFI is given as
\begin{equation}
\begin{array}{llll}
F_{\gamma}(t) \simeq \frac{1}{32}e^{-t \gamma_{F}}t^2.
\end{array} \label{Fishfeedlimit}
\end{equation}
with a rate $\gamma_{F}=\gamma /4$.
This smaller decay rate means the precision of parameter estimation is enhanced by the feedback. In Fig.~\ref{FisherA}, we plot the QFI with respect to time $t$. This figure shows that in the short time, the value of the QFI without feedback is larger than that with feedback. Fig.~\ref{FisherA} also displays that the QFI with the feedback decays slowly than that without the feedback in the long time. Moreover, the maximum value of the QFI is enhanced by the feedback. With the increase of the decay rate, the value of the QFI drops
considerably.

%%%%%%%%%%%%%%%%%%%%%%%%
%%%% Figure 4:
%%%%%%%%%%%%%%%%%%%%%%%%
\begin{figure}[!tb]
\centering
\includegraphics[width=2.0in]{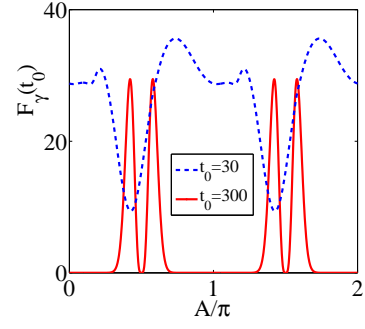}
\hspace{2.2cm}\caption{(Color online) The fixed time QFI $F_{\gamma}(t_{0})$
in terms of the feedback parameters $A$ with $\beta=0$.
The dashed (solid) line corresponds to the time $t_{0}=20$ ($t_{0}=300$), respectively.
The feedback is switched off at $A=0, \pi$.
The other parameters are $\gamma=0.1$. } \label{FisherB}
\end{figure}
%%%%%%%%%%%%%%%%%%%%%%%%
%%%%%%%%%%%%%%%%%%%%%%%%

%%%%%%%%%%%%%%%%%%%%%%%%
%%%% Figure 5:
%%%%%%%%%%%%%%%%%%%%%%%%
\begin{figure}[!tb]
\centering
\includegraphics[width=2.0in]{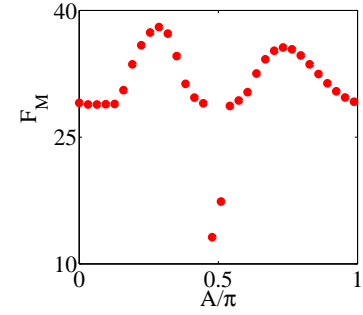}
\hspace{2.2cm}\caption{(Color online) The variation of of maximum value of the
QFI $F_{M}$ with respect to $A$. It is periodic function of $A$ with
a period $A=\pi$. The parameter is chosen as $\gamma=0.1$. } \label{FisherC2}
\end{figure}
%%%%%%%%%%%%%%%%%%%%%%%%
%%%%%%%%%%%%%%%%%%%%%%%%

For the general parameter, the QFI at the fixed time is plotted in Fig.~\ref{FisherB} in order to give further information of the feedback effect.
This figure is consistent with Fig.~\ref{FisherA}: in the long-time limit, the QFI is enhanced by the feedback. The peak structure in Fig.~\ref{FisherB} implies that the estimation precision can be strengthened by carefully tuning the feedback strength $A$ to an optimal value. We also find that
the QFI is a periodic function of $A$ with period $T=\pi$.

As the maximum value of QFI implies the largest precision to estimate a parameter, in Fig.~\ref{FisherC2} we plot $F_{M}$ as a function of $A$, with $F_{M}$ defined as
\begin{equation}
F_{M}=\max_{t}[\textit{F}_{\gamma}(t)].
\label{Fm}
\end{equation}
Without feedback ($A=0, \pi$), the maximum value of the QFI $F_{M}\approx 29.0$. At presence of the feedback, $F_{M}$ is enhanced, especially in the neighborhood of $A=\pi/3$ , $F_{M} \approx 38$. This figure also shows that near $A=\pi/2$, the quantum feedback paly a negative role for the QFI.

We have investigated the QFI for a specific initial state:
an equal-weight superposition pure state. Next,
we will study the behaviors of the QFI for other initial states,
especially a mixed one. For an initial mixed state $\rho_{0}=\varepsilon |e\rangle \langle e|+(1-\varepsilon) |g\rangle \langle g|$ ($\varepsilon \in [0, 1)$), the QFI for the general feedback parameters $A$ and $\beta$ is also very complicated. For the feedback value $A=\pi/3$, the QFI is obtained as
\begin{equation}
F_{\gamma}(t)= \frac{\eta_{1}(t)\varepsilon+\eta_{2}(t)\varepsilon^2+ \eta_{3}(t)\varepsilon^3 }{\eta_{4}(t)\varepsilon+\eta_{5}}
\label{FisherInitB}
\end{equation}
with
\begin{equation}
\begin{array}{llll}
\eta_{1}(t)= -e^{3t\gamma/2} t^2/2,~~
\eta_{2}(t)=( -12 e^{3 t\gamma/4} + 6 e^{t\gamma}) t^2,
\\
\eta_{3}(t)=6 e^{t \gamma/2} t^2,~~
\eta_{5}(t)= -8e^{7t\gamma/4}
\\
\eta_{4}(t)=8 ( e^{3 t\gamma/2} +3(e^{t\gamma} - 2 e^{5t\gamma/4} + e^{3 t\gamma}/2 )).
\end{array}
\end{equation}
In the limit $t\rightarrow 0$, this QFI reduces to
\begin{equation}
\begin{array}{llll}
F_{\gamma}(t)= \zeta(\varepsilon) t^2
\end{array}
\end{equation}
with an enhancing factor $\zeta(\varepsilon)=\frac{\varepsilon(-1-12\varepsilon+12\varepsilon^2)}{16(\varepsilon-1)}$
being a monotonically increasing function of $\varepsilon$. This means that the large probability in the excited state can enhance the precision of parameter estimation. In the long-time limit,
\begin{equation}
\begin{array}{llll}
F_{\gamma}(t)\approx \frac{1}{16} \varepsilon e^{-t\gamma/4}t^2,
\end{array}
\end{equation}
which also goes up with the increase of $\varepsilon$.

The effect of the initial state on the QFI is further studied
by the numerical simulation. We find that the maximum value of the QFI linearly (approximately) increases with the probability in the excited state $\varepsilon$. The physical reason of these results can be understood as follows: The information of $\gamma$ comes from the effective damping of the excited state.
The initial state with a high probability in the excited state should produce a larger QFI value. In absence of the feedback (i.e., $A=\pi$), the QFI is gained as $F_{\gamma}(t)=\frac{\varepsilon t^2}{e^{t \gamma}-\varepsilon}$, which has larger exponential decay than that of $A=\pi/3$ in the long-time limit. As a short summary, the QFI can be enhanced by the feedback, independent of the input state.

\subsection{The QFI of the steady state with driving}

In the previous subsection, we study the QFI with respect to the effective damping parameter $\gamma$ with switching off the external driving. The main result is the QFI can be enhanced by the feedback. However, in this case of no external driving, the QFI decays to zero exponentially in the long-time limit. Here we extend our study to the case of the external driving strength $\Omega \neq 0$. In this situation, the time-evolving density matrix of the qubit at any time can not be analytically given. However, one can obtain the non-zero steady state in the long-time limit (actually, in many realistic information processing, the steady state is more useful) as
\begin{equation}
\begin{array}{llll}
\rho_{s, 11}= \Omega ^2/ M, \ \ \rho_{s, 12}= N/M
\end{array}
\end{equation}
with
\begin{equation}
\begin{array}{llll}
N=\Omega [\Omega \sin(2A) \cos(\beta)-i\cos(A)^2\gamma],
\\
\\
M=\Omega \sin(2A)\gamma \sin(\beta)+2\Omega^2+\cos(A)^2\gamma ^2.
\end{array}
\end{equation}
We have checked that the QFIM of this steady state satisfies $\mathrm{Det}(\textbf{F}(g, \kappa))= 0$. The above expressions for the steady state is also complicated, then we investigate its property by considering specific parameters. In the strong-driving limit, i.e., $\Omega$ is much larger than the other parameters, the elements of the state reduce to $\rho_{s, 11}=1/2$ and $\rho_{s, 12}=0$, and the corresponding QFI with respect to $\gamma$ equals to zero.

%%%%%%%%%%%%%%%%%%%%%%%%
%%%% Figure 6:
%%%%%%%%%%%%%%%%%%%%%%%%
\begin{figure}[!tb]
\centering
\includegraphics[width=2.0in]{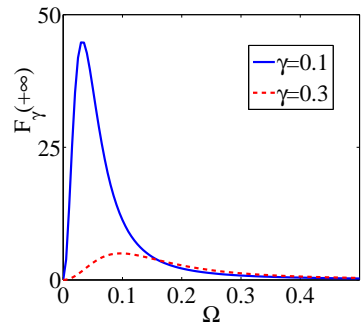}
\hspace{2.2cm}\caption{(Color online) The QFI of the steady state with respect to $\Omega$.
There is an optimal driving strength $\Omega_{m}$ to maximize the $F_{\gamma}(\infty)$.
The other parameters are $A=\pi/3$. All the relevant quantity is dimensionless, such as
$\Omega$ is scaled by $\Omega~\delta t$. } \label{FisherS}
\end{figure}
%%%%%%%%%%%%%%%%%%%%%%%%
%%%%%%%%%%%%%%%%%%%%%%%%

By setting $A=\pi/3$,
the elements of the density matrix translate to $\rho_{s, 11}= \Omega ^2/ \chi$,
$\rho_{s, 12}= \Omega (\sqrt{3}\Omega -i\gamma/2)/ \chi$, with $\chi=\frac{1}{2}\gamma^2+4\Omega^2$.
The QFI of this steady state with respect to $\gamma$ is obtained as
\begin{equation}
\begin{array}{llll}
F_{\gamma}=\frac{16\Omega^2(3\gamma^2 +\Omega^2)}
{ (3\gamma^2 +4\Omega^2 )(\gamma^2+8\Omega^2)^2}.
\end{array}
\end{equation}
It is obvious that without the driving ($\Omega=0$), the QFI of the steady
state equals to zero. In the limit $\Omega \rightarrow 0$, the QFI becomes
$F_{\gamma} \propto \Omega^2$, which goes up with the increase of $\Omega$.
Taking into consideration the fact that the QFI is close to zero in the large driving limit, there should have an optimal driving strength $\Omega_{m}$ to maximize the QFI. In Fig.~\ref{FisherS}, we show the variation of the QFI with respect to $\Omega$, which is consistent with the analytical analysis.
This figure also shows that with the increase of $\gamma$, the value of the QFI drops considerably, which implies the precision of estimation of $\gamma$ decreases.

\section{Conclusion}
In summary, in the system that the qubit couples to the cavity, we demonstrate that the quantum feedback can enhance the the precision of parameter estimation.
Quantum Fisher information is used as a measure to describe the parameter-estimation precision as one can use the observable composed by the eigenvectors of the symmetry logarithmic derivative to saturate the quantum Cram\'{e}r-Rao inequality. Different from the other quantum correlation
such as quantum entanglement and quantum steering,
the quantum Fisher information characterizes the sensitivity of a quantum state
with respect to the change of a certain parameter.
We find that in presence of the feedback, the decay rate of the quantum Fisher information is slowed down and the maximum value of the quantum Fisher information is also strengthened. In future, it's an interesting topic to explore the effects of a homodyne measurement scheme and the detection efficiency on the quantum Fisher information. In addition, the quantum Fisher information of multi-qubits \cite{Stevenson10} are another valuable subject.
\\
\\
\textit{Acknowledgments.}
We thank Profs. X. G. Wang, Y. Li and Dr. X. W. Xu for their helpful discussions.
This work is partially supported by the National Natural Science
Foundation of China (Grant Nos.~11365006), and the Innovation Team
Foundation of the Education Department of Guizhou Province under Grant No. [2014]35.

%%%%%%%%%%%%%%%%%%%%%%%%%%
%\newpage

%\end{CJK*}

\end{document}